\documentclass[notitlepage,11pt]{article}
\usepackage{hyperref}
\usepackage[hyphenbreaks]{breakurl}
\usepackage{graphicx} 
\usepackage{amssymb,amsfonts,amsmath}
\usepackage{txfonts}
\usepackage[]{caption} 
\usepackage{xcolor}              
\usepackage{setspace}            
\usepackage{authblk}             
\usepackage{fullpage}            
\usepackage{url}
\usepackage{lastpage,fancyhdr}
\usepackage{hyperref}
\usepackage[]{pdfpages}
\usepackage{footnote}
\makesavenoteenv{tabular}

\def\lett#1{(\textbf{#1})}

\def\SI{SI}

\def\lett#1{(\textbf{#1})}

\def\SI{SI}

\makeatletter
\newlength\figwidthone
\newlength\figwidthtwo
\makeatother

\title{Understanding the group dynamics and success of teams}

\author[1]{Michael Klug}
\author[1,2,3,$^*$]{James P.~Bagrow}
\affil[1]{\small Department of Mathematics \& Statistics, The University of Vermont}
\affil[2]{Vermont Complex Systems Center, The University of Vermont}
\affil[3]{Vermont Advanced Computing Core, The University of Vermont, Burlington, VT, USA}
\affil[$^*$]{To whom correspondence should be addressed. E-mail: james.bagrow@uvm.edu}

\begin{document}

\maketitle

\begin{abstract}
Complex problems often require coordinated group effort
and can consume significant resources, yet our understanding of how teams form and
succeed has been limited by a lack of large-scale, quantitative data.
We analyze activity traces and success levels for $\sim$150,000 self-organized,
online team projects.
While larger teams tend to be more successful, workload is
highly focused across the team, with only a few members performing
most work. 
%
%
We find that highly successful teams are significantly more focused than average
teams of the same size, that their members have worked on more diverse sets of
projects, and the members of highly successful teams are more likely to be core members or `leads' of
other teams.
The relations between team success and size, focus and especially team experience
 cannot be explained by confounding factors such as team age, external contributions from non-team members, nor by
group mechanisms such as social loafing. Taken together, these features point to organizational
principles that may maximize the success of collaborative endeavors.
\end{abstract}

\section{Introduction}

Massive datasets describing the activity patterns of large human
populations now provide researchers with rich opportunities to quantitatively
study human dynamics~\cite{lazer2009life,vespignani2012modelling}, including the
activities of groups or teams~\cite{dyer2009leadership,pentland2012new}.
New tools, including electronic sensor systems, can quantify team activity and
performance~\cite{choudhury2003sensing,pentland2012new}.
With the rise in prominence of network
science~\cite{barabasi2002linked,newman2010networks}, much
effort has gone into discovering meaningful groups within social
networks~\cite{wasserman1994social,sparrowe2001social,girvan2002community,watts2002identity,newman2003social,bird2008latent,ahn2010link,mucha2010community} and
quantifying their evolution~\cite{palla2007quantifying,mucha2010community}.
Teams are increasingly important in research and industrial
efforts~\cite{brooks1995mythical,faraj2000coordinating,rising2000scrum,dyer2009leadership,pentland2012new,milojevic2014principles,Montjoye:2014aa} and
small, coordinated groups are a significant component of modern human
conflict~\cite{clauset2007frequency,bohorquez2009common}.
There are many important dimensions along which teams should be studied, including
their size, how work is distributed among their members, and the differences and similarities
in the experiences and backgrounds of those team members.
Recently, there has been much debate on the ``group size hypothesis'', that
larger groups are more robust or perform better than smaller
ones~\cite{henrich2004demography,derex2013experimental,Andersson:2014aa,Derex:2014aa}.
Scholars of science have noted for decades that collaborative research teams
have been growing in size and
importance~\cite{price1963solla,pao1992global,hudson1996trends,milojevic2014principles}.
At the same time, however, social loafing, where individuals
apply less effort to a task when they are in a group than when
they are alone, may counterbalance the effectiveness of larger
teams~\cite{latane1979many,harkins1987social,karau1993social}.
Meanwhile, case studies show that
leadership~\cite{denis2001dynamics,dyer2009leadership,johnstone2011evolution,contractor2012topology}
and experience~\cite{katzenbach1993wisdom,delmar2006does} are key components of
successful team outcomes, while specialization and multitasking are important
but potentially error-prone mechanisms for dealing with complexity and cognitive
overload~\cite{lindbeck2000multitask,postrel2002islands}.
In all of these areas, large-scale, quantitative data can push the study of
teams forward.

Teams are important for modern software engineering tasks, and  researchers have long studied the digital traces of open source
software projects to better quantify and understand how teams work on software
projects~\cite{crowston2012free,Scholtes2015}. 
Researchers have investigated estimators of work activity or effort based on
edit volume, such as different ways to count the number of changes made to a software's 
source code~\cite{albrecht1983software,rosenberg1997some,koch2002effort,alali2008s}.  
Various dimensions of success of software projects, such as popularity, timeliness of bug fixes, or 
other quality measures have been studied~\cite{crowston2006information,subramaniam2009determinants,ghapanchi2011taxonomy}.
Successful open source software projects show a layered structure of primary or core contributors surrounded
by lesser, secondary contributors~\cite{crowston2005social}.
At the same time, much work is focused on case studies~\cite{mockus2002two,koch2002effort} of small numbers of highly successful, large projects~\cite{crowston2012free}.
Considering these studies alone runs the risk of survivorship bias or other selection biases, 
so large-scale studies of large quantities of teams are important complements to these works.

Users of the GitHub web platform can form teams to work on real-world projects,
primarily software development but also music, literature, design work, and more. 
A number of important scientific computing resources are now
developed through GitHub, including astronomical software, genetic sequencing tools, and key components of the
Compact Muon Solenoid experiment's data pipeline.\footnote{For examples, 
see %
    \url{https://github.com/showcases/science}.} 
A ``GitHub for science'' initiative has been launched\footnote{See
    \url{https://github.com/blog/1840-improving-github-for-science}.}
and GitHub is becoming the dominant service for open scientific development.

GitHub provides rich public data on team activities, including when new teams
form, when members join existing teams, and when a team's project is updated.
%
%
GitHub also provides social media tools for the discovery of interesting
projects.
Users who see the work of a team can choose to flag it as interesting to them by ``starring'' it.
The number of these ``stargazers'' $S$ allows us to quantify one aspect of the \textbf{success} of the team, in a
manner analogous to the use of citations of research
literature as a proxy for ``impact''~\cite{wang2013quantifying}. 
Of course, as with bibliometric impact, one should be cautious and not consider success to be a perfectly accurate measure of \emph{quality}, something that is far more difficult to objectively quantify. Instead this is a measure of popularity as would be other statistics such as web traffic, number of downloads, and so forth~\cite{crowston2006information}.
%

%
%

%
%
%
%
%
%
%
%
%
%
%
%
%
%
%
%
%
%
%
%
%
%
%
%
%
%

In this study, we analyze the memberships and activities of approximately
150,000 teams, as they perform real-world tasks, to uncover the
blend of features that relate to success.
To the best of our knowledge this is the largest study of real-world team
success to date.
We present results that demonstrate (i) how teams distribute or focus work
activity across their members, (ii) the mixture of experiential diversity
and collective leadership roles in teams, and (iii) how successful teams are
different from other teams while accounting for confounds such as team size.

The rest of this paper is organized as follows:
In Sec.~\ref{sec:methods} we describe our GitHub dataset; give definitions of a team, team
success, and  work activity/focus of a team member; and introduce metrics to
measure various aspects of the experience and experiential diversity of a team's members.
In Sec.~\ref{sec:results} we present our results relating these measures to team success.
In Sec.~\ref{subsec:combined} we present statistical tests on linear regression
models of team features to control for potential confounds between team features
and team success.
Lastly, we conclude with a discussion in Sec.~\ref{sec:discuss}.

\section{Methods and Materials}
\label{sec:methods}
    
\subsection*{Dataset and team selection} 

Public GitHub data covering 1 January 2013 to 1 April 2014 was collected from
githubarchive.org in April 2014. In their own words, ``GitHub Archive is a
project to record the public GitHub timeline, archive it, and make it easily
accessible for further analysis''.
These activity traces contain approximately
110M unique events, including when users create, join, or update projects.
Projects on GitHub are called ``repositories''. For this work we define a
\textbf{team} as the set of users who can directly update (``push to'') a
repository. These users constitute the \textbf{primary} team members as they have
either created the project or
been granted autonomy to work on the project. The number of team members was
denoted by $M$. Activity or workload $W$ was estimated from the number of pushes.
A push is a bundle of code updates (known as commits), however most pushes contain
only a single commit (see \SI{}; see also ref.~\cite{alali2008s}). As with all studies measuring worker effort from lines-of-code
metrics, this is an imperfect measure as the complexity of a unit of work does not generally
map to the quantity of edits.
Users on GitHub can bookmark projects they find interesting. This is called
``stargazing''. We take the maximum number of stargazers for a team as its
measure of \textbf{success} $S$. This is a popularity measure of success, however the choice
to bookmark a project does imply it offers some value to the user.
To avoid abandoned projects, studied teams have at least one stargazer ($S>0$)
and at least two updates per month on average within the githubarchive data. 
%
These selection criteria leave $N = 151,542$ teams. 
We also collect the time of creation on GitHub for each team project.
This is useful for measuring confounds: for example, older teams may tend to both have
more members and have more opportunities to increase success. Of the teams
studied, 67.8\% were formed within our data window. Beyond considering team age as a potential confounder, we do not study temporal dynamics such as team formation in this work.
A small number of studied teams
(1.08\%) have more than ten primary members ($M>10$); those teams were not shown in figures, but they were
present in all statistical analyses.
Lastly, to ensure our results are not due to outliers, in some analyses we excluded teams above the 99th percentile of $S$. Despite a strong skew in the distribution of $S$, these highly popular teams account for only 2.54\% of the total work activity of the teams considered in this study (2.27\% when considering teams with $M\leq10$ members). 

\paragraph{Secondary Team} GitHub provides a mechanism for external, non-team contributors
to propose work that team members can then choose to use or not. These proposals
are called pull requests. (Other mechanisms, such as discussions about issues, are also available to non-team contributors.)
These secondary or external team contributors are not the focus of this
work and have already been well studied by OSS researchers~\cite{crowston2012free}. However, 
it is important to ensure
that they do not act as confounding factors for our results, since more successful teams will tend to have more secondary contributions than other teams. So we measure for each team $M_\mathrm{ext}$,
the number of unique users who submit at least one pull request, and $W_\mathrm{ext}$ the number of pull requests.
We will include these measures in our combined regression models.
Despite their visibility in GitHub, pull requests are rare~\cite{kalliamvakou2014promises}; in our data, 57.7\% of teams
we study have $W_\mathrm{ext}=0$, and when present pull requests are greatly
outnumbered by pushes on average: $\left< W / W_\mathrm{ext} \mid W_\mathrm{ext}
> 0 \right> = 42.3$ (median $16.0$), averaged over all teams with at least one
pull request.

\subsection*{Effective team size}
The number of team members, $M$, does not fully represent the size of a team
since the distribution of work may be highly skewed across team members. To
capture the \textbf{effective team size} $m$, accounting for the relative
contribution levels of members, we use $m = 2^{H}$, where $H = -\sum_{i=1}^M
f_i \log_2 f_i$, and $f_i = w_{i}/W$ is the fraction of work performed by team
member $i$.
This gives $m=M$ when all $f_i = 1/M$, as expected.  This simple, entropic
measure is known as perplexity in linguistics and is closely related to species
diversity indices used in ecology and the Herfindahl-Hirschman Index used in
economics.

\subsection*{Experience, diversity, and leads}
Denote with $R_i$ the set of projects that user $i$ works on (has pushed to).  (Projects in $R_i$
need at least twice-monthly updates on average, as before, but may have $S=0$ so
as to better capture $i$'s  background, not just successful projects.) We estimate
the \textbf{experience} $E$ of a team of size $M$ as 
\begin{equation*}
    E = \frac{1}{M}\sum_i \left| R_i \right|-1
\end{equation*}
and the experiential \textbf{diversity} $D$ as
\begin{equation*}
    D = \frac{\left| \bigcup_i R_i \right|}{\sum_i \left| R_i\right|},
\end{equation*}
where the sums and union run over the $M$ members of the team. Note that 
$D \in \left[1/M,1\right)$.
Experience measures the quantity of projects the team works on while diversity measures
how many or how few projects the team members have in common, the goal being to capture
how often the team has worked together.
Lastly, someone is a \textbf{lead} when, for at least one project they work on, they
contribute more work to that project than any other member. A non-lead member of
team $j$ may be the lead of project $k\neq j$. The number of leads $L_k$ in team
$k$ of size $M_{k}$ is:
\begin{equation*}
    L_k = \sum_{i=1}^{M_k} \min\left(\sum_j L_{ij}, 1\right),
\end{equation*}
where $L_{ij} = 1$ if user $i$ is the lead of team $j$, and zero otherwise. The
first sum runs over the $M_k$ members of team $k$, the second runs over all
projects $j$.
Of course, the larger the team the more potential leads it may contain so when
studying the effects of leads on team success we only compare teams of the same
size (comparing $L$ while holding $M$ fixed). Otherwise, $E$ and $D$ already
account for team size.

\section{Results}
\label{sec:results}

\begin{figure*}[t]
\begin{center}
    \includegraphics[width=\figwidthone]{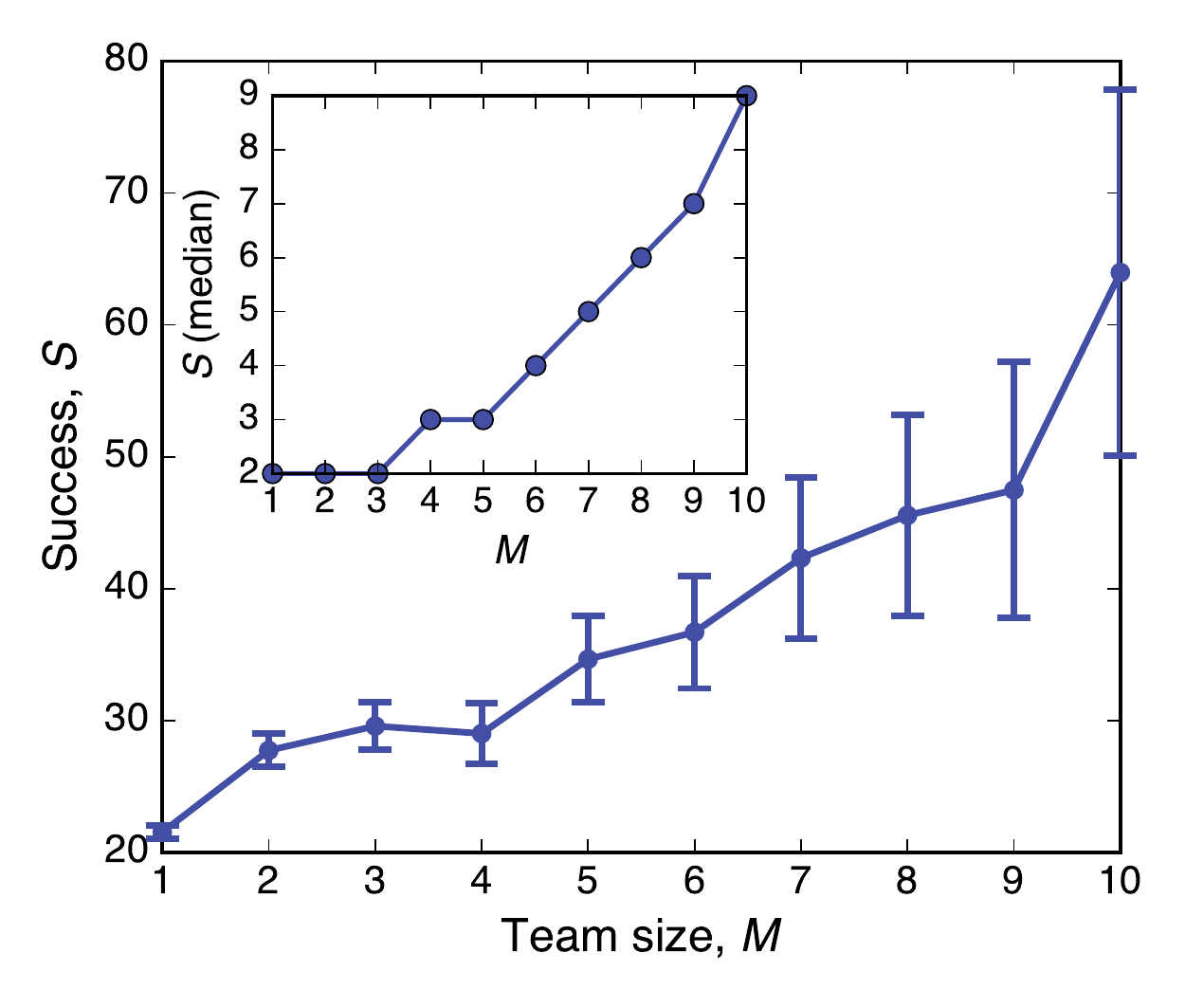}
\caption{
Larger teams have significantly more success on average, with a 300\% increase
in $S$ as $M$ goes from 1 to 10. This correlation may be due to more team
members driving project success or success may act as a mechanism to recruit
team members.
Errorbars here and throughout denote $\pm 1.96$ s.e.
(\textbf{Inset}) Using the median instead of the mean shows that this trend is not due to outliers.
\label{fig:introToData}
}
\end{center}
\end{figure*}

We began our analysis by measuring team success $S$ as a function of team
size $M$, the number of primary contributors to the team's project. Since $S$
is, at least partially, a popularity measure, we expect larger teams to also be
more successful. Indeed, there was a positive and significant relationship
($p<10^{-10}$, rank correlation $\rho = 0.0845$) between the size of a team and
its success, with 300\% greater success on average for teams of size $M=10$
compared to solos with $M=1$ (Fig.~\ref{fig:introToData}). This strong trend
holds for the median success as well (inset).
While this observed trend was highly significant, the rank correlation $\rho$
indicates that there remains considerable variation in $S$ that is not captured
by team size alone.

Our next analysis reveals an important relationship between team focus and
success. 
Unlike bibliographic studies, where teams can only be quantified as the listed
coauthors of a paper, the data here allow us to measure the intrinsic work or
volume of contributions from each team member to the project.
For each team we measured the contribution $w_r$ of a member to the team's
ongoing project, how many times that member updated the project (see Methods).
Team members were ranked by contribution, so $w_1$ counts the work of the member
who contributed the most, $w_2$ the second heaviest contributor, and so
forth. 
The total work of a team is $W = \sum_{r=1}^{M}w_r$.

\begin{figure}[t!]
\centering
    \includegraphics[width=\figwidthtwo]{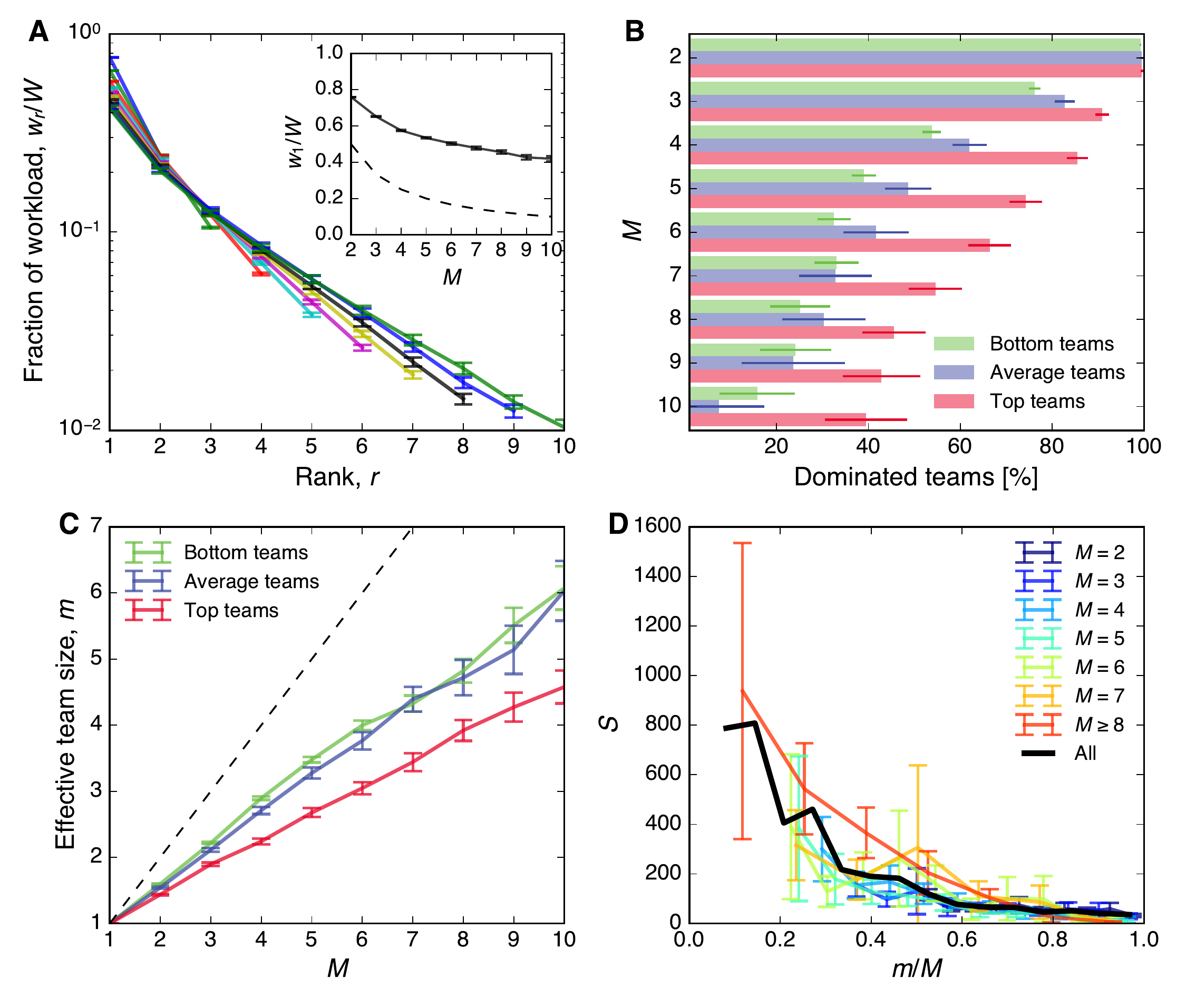}
\caption{Teams are focused, and top teams are more focused than other teams of
the same size. %
\lett{A} %
The average fraction of work $w_r /W$ performed by the $r$-th most active
member, where $W$ is the total work of the team, for different size teams.
Larger teams perform more work overall, but the majority of work is always done
by a small subset of the $M$ members (note the logarithmic axis). 
Inset: The fraction of work performed by the most active team member is always
high, often larger than half the total. The dashed line indicates the lower
bound of uniform work distribution,  $w_r/W = 1/M$.
\lett{B} %
A team is \textbf{dominated} when the most active member does more work than all other
members combined. Top teams are significantly
more likely to be dominated than either average teams or bottom teams for all $M>2$.
(\textit{Top team}: above the 90th percentile in $S$; \textit{average team}: greater than the 40th percentile of $S$ and less than or equal to the 60th percentile of $S$; \textit{bottom team}: at or below the 10th percentile of $S$.)
\lett{C} %
The effective team size $m$ (see Methods), a measure that accounts for the
skewed distribution of work in panel A, is significantly smaller than $M$. Moreover,
top teams are significantly more focused, having smaller effective sizes, than
average or bottom teams at all sizes $M>1$. This includes the case $M=2$, which did not show a significant difference in panel B. The dashed line denotes the upper bound $m=M$.
\lett{D} %
Success is universally higher for teams with smaller $m/M$, independent of
$M$, further supporting the importance of focused workloads. The solid lines
indicates the average trend for all teams $2\leq M \leq 10$.
These results are not due to outliers in $S$; see \SI{}.
\label{fig:teamsAreFocused} }
\end{figure}

We found that the distribution of work over team members showed significant
skew, with $w_1$ often more than 2--3 times greater than $w_2$
(Fig.~\ref{fig:teamsAreFocused}A and \SI{}). This means
that the workloads of projects are predominantly carried by a handful of team
members, or even just a single person.
Larger teams perform more total work, and the heaviest contributor carries much of
that effort: the inset of Fig.~\ref{fig:teamsAreFocused}A %
shows that $w_{1}/W$, the fraction of work carried by the rank one member, falls slowly with
team size, and is typically far removed from the lower bound of equal work among all team
members.
See \SI{} for more details.
This result is in line with prior studies \cite{mockus2002two}, supporting the 
plausibility of our definition of a team and our use of pushes to measure work.

This focus in work activity indicates that the majority of the team serves as a
support system for a core set of members. Does this
arrangement play a role in whether or not teams are successful?
We investigated this in several ways. First, we asked whether or not a team was
\textbf{dominated}, meaning that the lead member contributed more work than all
other members combined ($w_1/W > 1/2$). Highly successful ``top'' teams,
those in the top 10\% of the success distribution, were significantly more likely
to be dominated than average teams, those in the middle 20\% of $S$, or ``bottom'' teams, those in bottom 10\% of the $S$.
(Fig.~\ref{fig:teamsAreFocused}B).

Can this result be due to a confounding effect from success? More
successful projects will tend to have more external contributors, for example,
which can change the distribution of work. For example, in one scenario a team
member may be a ``community manager'' merging in large numbers of external contributions
from non-team members. To test this we examined only the 57.7\% of teams
that had no external contributions ($W_\mathrm{ext} = 0$) and tested among only
those teams whether dominated teams were more successful than non-dominated
teams. Within this subset of teams, dominated teams had significantly higher $S$
than non-dominated teams (Mann-Whitney U test with continuity correction, $p <
10^{-8}$). The Mann-Whitney U test (MWU) is non-parametric, using ranks of (in this case) $S$ 
to mitigate the effects of skewed data, and does not assume normality.
We conclude from this that external contributions do not fully explain the
relationship between workload focus and team success.

Next, we moved beyond the effects of the heaviest contributor by performing the
following analysis. For each team we computed its \textbf{effective} team size
$m$, directly accounting for the skew in workload (see Methods for full
details). 
This effective size can be roughly thought of as the average number of unique
contributors per unit time and need not be a whole number. For example, a team
of size $M=2$ where both members contribute equally will have effective size
$m=2$, but if one member is responsible for 95\% of the work the team would have
$m \approx 1.22$.
Note that $M$ and $m$ are positively correlated ($\rho = 0.985$).

Figure~\ref{fig:teamsAreFocused}C shows that (i) all teams are effectively much
smaller than their total size would indicate, for all sizes $M>1$, and (ii) top
teams are significantly smaller in effective size (and therefore more focused in
their work distribution) than average or bottom teams with the same $M$. 
Further, success is significantly, negatively correlated with $m$, for
all $M$ (Fig.~\ref{fig:teamsAreFocused}D). More focused teams have significantly
more success than less focused teams of the same size, regardless of total team size.
%

Further analyses revealed the importance of team composition and its role in
team success. 
%

Team members do not perform their work in a vacuum, they each bring experiences
from their other work. Often members of a team will work on other projects. We
investigated these facets of a team's composition by exploring (i) how many
projects the team's members have worked on, (ii) how diverse are the other projects
 (do the team members have many or few other projects in
common), and (iii) how many team members were ``leads'' of other projects.

An estimate of experience, $E$, the average number of other projects that team members have
worked on (see Methods), was significantly related to success. However, the trend
was not particularly strong (see \SI{}) and, as we later show via combined
modeling efforts, this relationship with success was entirely explainable by the
teams' other measurable quantities.

It may be that the  volume of experience does not contribute much to the success
of a team, but this seems to contradict previous studies on the importance of
experience and wisdom~\cite{katzenbach1993wisdom,delmar2006does}. To
investigate, we turned to a different facet of a team's composition, the
diversity of the team's background. 
Successful teams may tend to be comprised of members who have frequently worked
together on the same projects in the past, perhaps developing an experiential
shorthand. Conversely, successful teams may instead have multiple distinct
viewpoints, solving challenges with a multi-disciplinary
perspective~\cite{horwitz2007effects}.

\begin{figure}[t]
\centering
    {\includegraphics[width=\figwidthone]{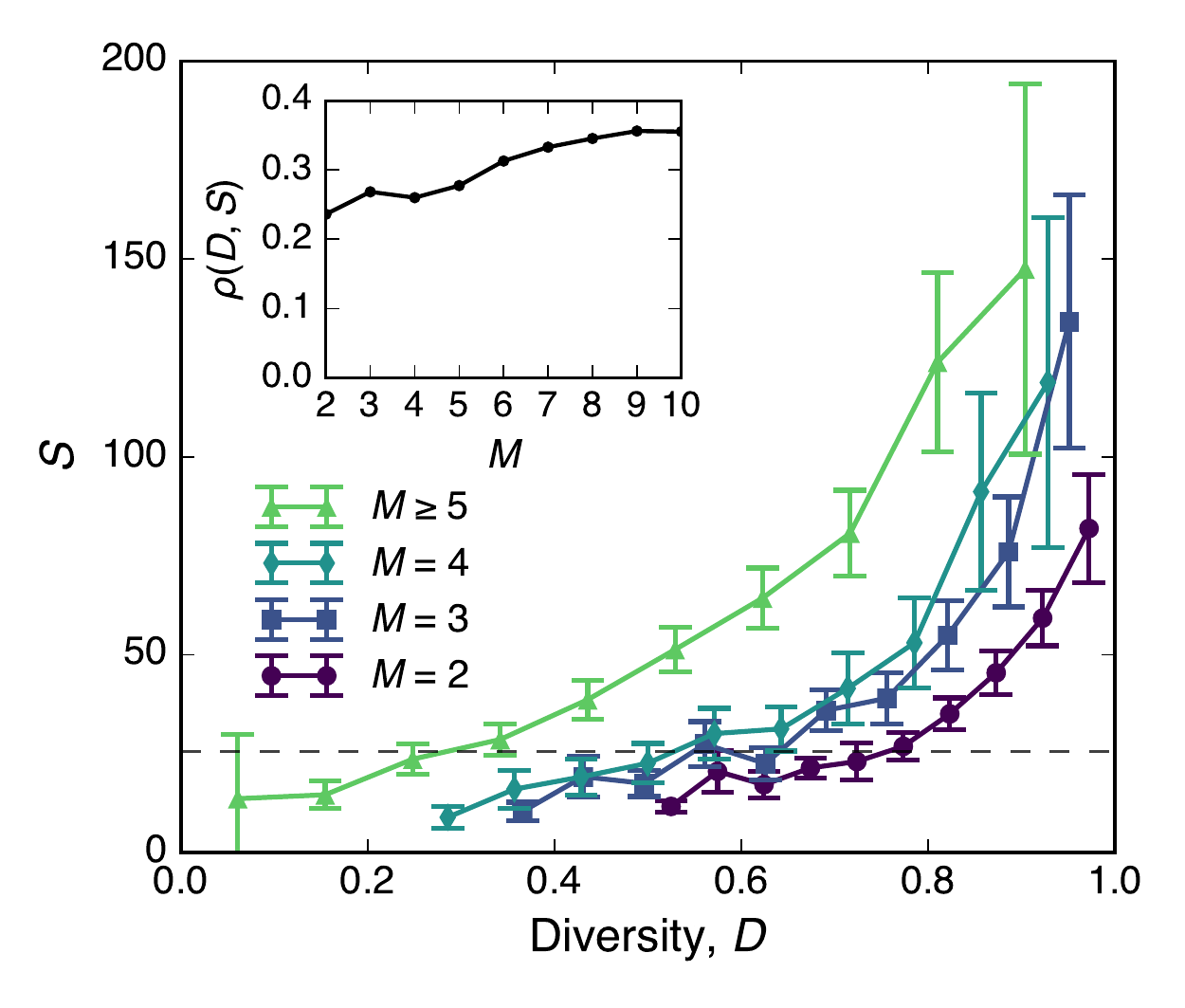}}
\caption{Teams whose members belong to more diverse sets of other teams tend to
    be more successful, regardless of team size. The dashed line denotes the average success of all teams. (Inset) The rank correlation
    $\rho$ between diversity and success grows with team size. 
Teams above the 99th percentile in $S$ were excluded to ensure the trend is not due to outliers.
\label{fig:teamComposition}}
\end{figure}

To estimate the distinctness of team member backgrounds, the diversity $D$ was
measured as the fraction of projects that team members have worked on that are
unique (see Methods). Diversity is low when all $M$ members have worked on
the same projects together ($D = 1/M$), but $D$ grows closer to $1$ as their
backgrounds become increasingly diverse.
A high team diversity was significantly correlated with success, regardless
of team size (Fig.~\ref{fig:teamComposition}). Even small teams seem to have
benefited greatly from diversity: high-$D$ duos averaged nearly \textbf{eight times} the
success of low-$D$ duos. The relationship between $D$ and $S$ was even stronger
for larger teams (Fig.~\ref{fig:teamComposition} inset), implying that larger
teams can more effectively translate this diversity into success.
Even if the raw volume of experience a team has does not play a significant role
in the team's success, the diversity of that experience was significantly correlated
with team success.
See also our combined modeling efforts.

Considerable attention has been paid recently to collective leadership, where
decision-making structures emerge from the mass of the group instead of being
imposed via a top-down
hierarchy~\cite{denis2001dynamics,contractor2012topology}.  The open
collaborations studied here have the potential to display collective leadership
due to their volunteer-driven, self-organized nature.
The heaviest contributor to a team is most likely to occupy such a leadership
role.
Further, since teams overlap, a secondary member of one team may be the
``lead,'' or heaviest contributor to another. This poses an interesting
question: Even though teams are heavily focused, are teams more successful when
they contain many leads, or few? A team with many leads will bring considerable
experience, but most of its members may also be unable to dedicate their full attention to the team.

\begin{figure}[t]
\centering
    {\includegraphics[width=0.5\figwidthtwo,trim=100 25 40 55,clip=true]{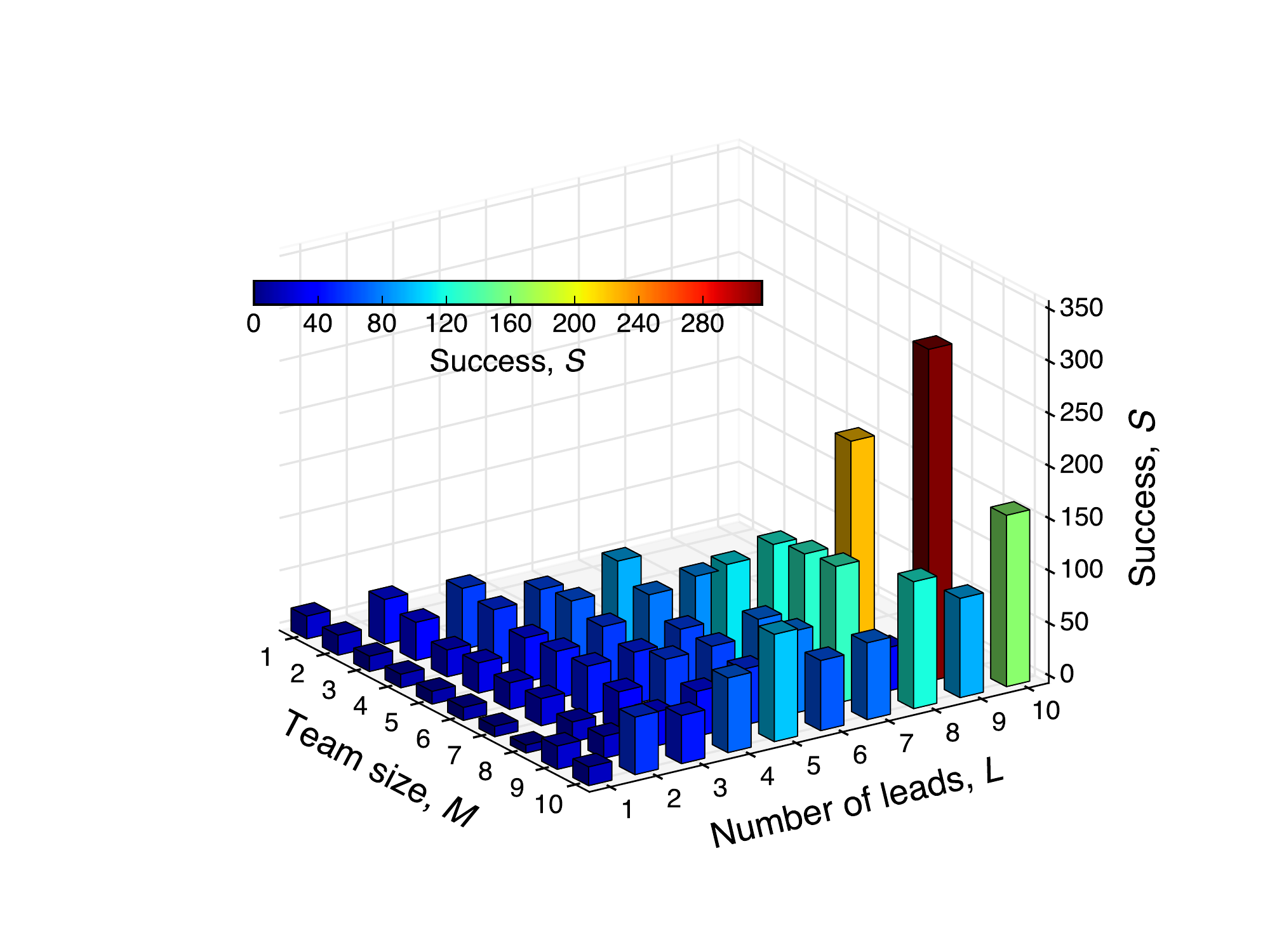}}
    \includegraphics[width=0.5\figwidthtwo]{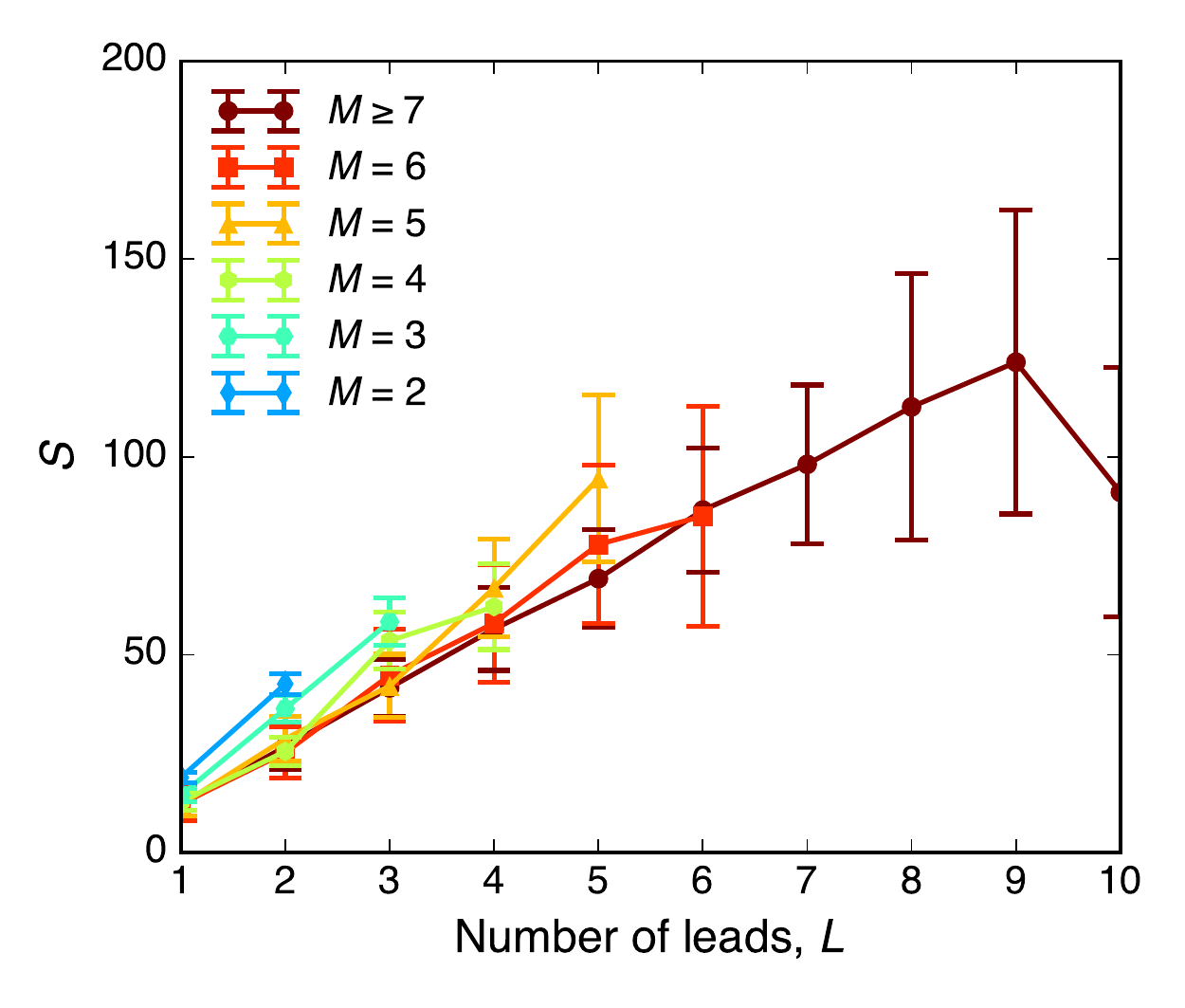}
\caption{Teams with more leads have higher success than teams of the same size
    with fewer leads.
    A lead is someone who contributes more work to at least one team he or she
    belongs to than any other members of that team.
    Outliers in $S$ were removed as before.
\label{fig:teamsHaveLeads}}
\end{figure}

To answer this, we measured $L$, the number of team members who are the lead of
at least one project ($1 \leq L \leq M$, see Methods), and found that teams with
many leads have significantly higher success than teams \emph{of the same size} with
fewer leads (Fig.~\ref{fig:teamsHaveLeads}).
Only one team member can be the primary contributor to the team, so a team can only have many leads if  the other members have focused their work activity on other projects. 
Team members that are focused on other projects can potentially only provide limited support, yet
successful teams tend to arrange their members in exactly this
fashion. 
Of course, the strong focus in work activity (Fig.~\ref{fig:teamsAreFocused}) is
likely interrelated with these observations. However, we will soon show that
both remain significantly related to success in combined models.

Expanding on this observation, Table~\ref{tbl:teamSizeLeadsImpact} illustrates
the extreme case of teams of size $M$ with a single lead ($L=1$) compared with teams of the same size
comprised entirely of leads ($L=M$). The latter always displayed significantly
higher success than the former (MWU test, see table), independent of team size,
underscoring the correlations displayed in Fig.~\ref{fig:teamsHaveLeads}. Often
the difference was massive: teams of size $M=7$, for example, averaged more than
1200\% higher success when $L=7$ than when $L=1$.  

\begin{table}[t]
\begin{minipage}{1\textwidth}
    \centering
\caption{Teams composed entirely of leads ($L=M$) are significantly more
    successful (MWU test on $S$) than teams of the same size with one lead ($L=1$), regardless of team
    size $M$.Teams above the 99th percentile in $S$ were excluded to ensure
    the differences were not due to outliers.
\label{tbl:teamSizeLeadsImpact} 
}
\begin{tabular}{llllllll}
\vrule depth -2pt width 0pt
&&\multicolumn2c{Num.\ teams $N$} && \multicolumn2c{Mean success $S$} &  \\ \cline{3-4} \cline{6-7}
\noalign{\vspace{2pt}}
$M$ && $L=1$ & $L=M$  && $L=1$ & $L=M$ & \multicolumn1c{MWU $p$-value} \\
\hline
$2$   &&  $14823$ & $8894$ &&   $18.9$ &  $42.5$  & $< 10^{-213}$\\  
$3$   &&  $6171$  & $2261$ &&   $14.5$ &  $58.3$  & $< 10^{-210}$\\  
$4$   &&  $3063$  & $717$  &&   $12.8$ &  $62.1$  & $< 10^{-112}$\\  
$5$   &&  $1489$  & $289$  &&   $12.1$ &  $94.5$  & $< 10^{-55 }$\\  
$6$   &&  $740$   & $124$  &&   $12.3$ &  $85.0$  & $< 10^{-36 }$\\  
$7$   &&  $350$   & $46$   &&   $9.8$  &  $120.5$ & $< 10^{-15 }$\\  
$8$   &&  $179$   & $19$\footnote{When $M\geq8$, the number of teams with $L=M$ is
    too small ($N<20$) for us to reasonably conclude the difference in $S$ is
    significant, despite the small $p$-values.}  &&  $7.5$  &  $224.1$ & $< 10^{-8  }$ \\
$9$   &&  $125$   & $9$    &&   $22.2$ &  $316.8$ & $< 0.008    $\\ 
$10$  &&  $66$    & $6$    &&   $17.8$ &  $163.5$ & $< 0.005    $\\
\end{tabular}
\end{minipage}
\end{table}

These results on team composition cannot be easily explained as a confound with success or secondary contributions as
they study specific features and projects of the individuals who comprise a team, those features are not related
to the successes of other projects an individual may work on, and they strictly control for total team size $M$
(for example, we only compare teams with different values of $L$ when they have the same value $M$).
These results further amplify our findings on
team focus, and augment important existing
research~\cite{katzenbach1993wisdom,dyer2009leadership,pentland2012new,contractor2012topology,horwitz2007effects}.

Taken together, our results demonstrate that successful teams tend to be
focused (Fig.~\ref{fig:teamsAreFocused}), successful teams tend to be experientially diverse (Fig.~\ref{fig:teamComposition}), and successful teams tend to have many leads (Fig.~\ref{fig:teamsHaveLeads}). 
We have found that teams tend to do best when optimized along all three of
these dimensions.  Of course, it is necessary to explore the joint effects of
quantities, to see if one relationship can be explained by another, which we will do with multivariate statistical models.

\section{Combined models and confounds}
\label{subsec:combined}
One important aspect of the individual team
measurements is that they do not exist in isolation. For example, successful
teams also have high work activity (high $W$). This can correlate with effective
team size $m$ since the potential inequality between team members can grow as
their total activity grows. In other words, we need to see how our team measures
relate to success \emph{together}.

To understand the relative effects of these team composition measures, we
fitted a linear regression model of success as a function
of all explored measures. See Table~\ref{tab:regressionResults}.  Not
only did this regression allow us to determine whether a variable was
significant or if it was confounded by the other measures, but the
coefficients (on the standardized variables) let us measure the relative
strengths of each variable. 
We also included the age of a project $T$ (measured as the time difference between the recorded creation time of the project and the end of our data window; see Methods) as this may be a potential confounding
factor as well (older projects have had more time to gain members and to gain
success).


\begin{table}[t]
\begin{minipage}{1\textwidth}
    \centering
\caption{OLS regression model on team success, 
$S = \alpha + \beta_M M + \beta_m m + \beta_W W + \beta_E E + \beta_D D + \beta_L L + \beta_T T$
Outliers (above the 99th percentile in $S$) were filtered out to ensure they do not skew the model.
\label{tab:regressionResults}}
\begin{tabular}{rr@{\hskip 1ex}cl}
\multicolumn1c{Variable $x$} &  \multicolumn2c{Coefficient $\beta_x$\footnote{Variables are standardized for comparison such that a coefficient $\beta_x$ implies that  increasing a variable $x$ by one standard deviation $\sigma_x$ corresponds to a $\beta_x \sigma_S$ increase in $S$, holding other variables fixed.}}       &   \multicolumn1c{$p$} \\
\hline{}
constant $\alpha$    &  $1.351 \times 10^{-14}$ &$\pm$ $0.004951$ & $1$          \\ %
Team size,       $M$ &                 $0.0848$ &$\pm$ $0.013963$ & $< 10^{-31}$ \\ %
Eff.\ team size, $m$ &                $-0.0989$ &$\pm$ $0.012140$ & $< 10^{-56}$ \\ %
Total work,      $W$ &                 $0.0323$ &$\pm$ $0.004997$ & $< 10^{-35}$ \\ %
Experience,      $E$ &              $0.0004068$ &$\pm$ $0.004985$ & $0.8729$     \\ %
Diversity,       $D$ &                $0.04099$ &$\pm$ $0.006357$ & $< 10^{-35}$ \\ %
Num.\ of leads,  $L$ &                 $0.1388$ &$\pm$ $0.006921$ & $0$          \\ %
Age,             $T$ &                 $0.1273$ &$\pm$ $0.005014$ & $0$
\end{tabular}
\end{minipage}
\end{table}

Examining the regression coefficients showed that the number of leads $L$ was the variable
most strongly correlated with team success.
Team age $T$, effective team size $m$, and team size $M$ play the strongest
roles after $L$ in team success, and all three were also significant in the presence of the
other variables.
The coefficient on $m$ was negative while for $M$ it was positive, further
underscoring our result that, while teams should be big, they effectively should
be small.
Next, the diversity $D$ of the team, followed by the total work $W$ done on the
project, were also significant measures related to success. Finally, overall
team experience $E$ was not significant in this model ($p > 0.1$). We conclude
that, while $S$ and $E$ are correlated by themselves, any effects of $E$ are
 explained by the other quantities.

What about secondary contributions, those activities made by individuals outside
the primary team? We already performed one test showing that dominated teams are more successful than non-dominated teams even when there are no secondary contributions. Continuing along these lines, we augmented this linear model with two more dependent variables,
$M_\mathrm{ext}$ and $W_\mathrm{ext}$. Regressing on this expanded model (see
\SI{} for details) did not change the significance of any coefficients at the $p=0.05$ level; $E$
remained insignificant ($p>0.1$). Both new variables were significant ($p < 0.05$).
Note that there were no multicollinearity effects in either regression model (condition
numbers < 10). We conclude that secondary contributions cannot alone explain
the observations relating team focus, experience, and lead number to team
success.

\section{Discussion}
\label{sec:discuss}

There has been considerable debate concerning the benefits of specialization
compared to diversity in the workplace and other
sectors~\cite{lindbeck2000multitask}.
Our discoveries here show that a high-success team forms a diverse support system
for a specialist core, indicating that both specialization and diversity
contribute to innovation and success.
Team members should be both specialists, acting as the lead contributor to a
team, and generalists, offering ancillary support for teams led by another
member.
This has implications when organizations are designing teams and wish to
maximize their success, at least as success was measured in these data.
Teams tend to do best on average when they maximize $M$
(Fig.~\ref{fig:introToData}B) while minimizing $m$
(Fig.~\ref{fig:teamsAreFocused}D) and maximizing $D$
(Fig.~\ref{fig:teamComposition}) and $L$ (Fig.~\ref{fig:teamsHaveLeads}).

Of course, some tasks are too large for a single person or small team to handle,
necessitating the need for mega-teams of hundreds or even thousands of members.
Our results imply that such teams may be most effective when broken down into
large numbers of small, overlapping groups, where all individuals belong to a
few teams and are the lead of at least one.  Doing so will help maximize the
experiential diversity of each sub-team, while ensuring each team has someone
``in charge''.
An important open question is what are the best ways to design such pervasively
overlapping groups~\cite{ahn2010link}, a task that may be project- or
domain-specific but which is worth further exploration.

The negative relationship between effective team size $m$ and success $S$ (as
well as the significantly higher presence of dominated teams among high success
teams) further belies the myth of multitasking~\cite{lindbeck2000multitask} and
supports the ``surgical team'' arguments of Brooks~\cite{brooks1995mythical}.
Focused work activity, often by even a single person, is a hallmark of
successful teams. This focus both limits the cognitive costs of task switching,
and lowers communication and coordination barriers, since so much work is being
accomplished by one or only a few individuals. We have provided statistical tests
demonstrating that the relationship between focus and success cannot be due to 
secondary/external team contributions alone.

Work focus could possibly be explained by \textbf{social loafing}
where individual members of a group contribute less effort as part of the group than they would alone, yet loafing does not explain 
the correlation between, e.g., leads and success (Fig.~\ref{fig:teamsHaveLeads}).
Likewise, our team composition results on group experience, experiential diversity, and the number of leads
cannot be easily explained as a confound with success or secondary contributions:
they study specific features of the individuals who comprise a team, those features are not related
to the successes of other projects an individual may work on, and they strictly control for total team size $M$
(except for the number of leads $L$, so for that measure we only compared teams with the same $M$).
The measures we used for external team contributions, $M_\mathrm{ext}$ and $W_\mathrm{ext}$, may be considered measures of success themselves,
and studying or even predicting their levels from team features may prove a fruitful avenue of future work.

Lastly, there are two remaining caveats worth mentioning. We do not specifically control for automatically
mirrored repositories (where a computer script copies updates to GitHub). Accurately detecting such projects at scale is a challenge beyond
the scope of this work. However, we expect most will either be filtered out by our existing selection
criteria or else they will likely only have a single (automated) user that only does the copying.
The second concern is work done outside of GitHub or, more generally, mismatched assignments between usernames and their work.
This is also challenging to fully address (one issue is that the underlying git repository system does not authenticate users).
We acknowledge this concern for our workload focus results, but even it  cannot explain the significant trends we observed on team composition such as the density of leads. Noise due to improperly recorded or ``out-of-band'' work has in principle affected all
quantitative studies of online software repositories.

\section*{Data accessibility}

All data analyzed are made publicly available by the GitHub Archive Project (\url{https://www.githubarchive.org}).

\section*{Competing interests}

We have no competing interests.

\section*{Authors' contributions}

MK participated in data collection and data analysis, and helped draft the manuscript; JB conceived of the study, designed the study, carried out data collection and analysis, and drafted the manuscript. All authors gave final approval for publication.

\section*{Acknowledgments}
    We thank Josh Bongard, Brian Tivnan, Paul Hines, Michael Szell, and
    Albert-L\'aszl\'o Barab\'asi for useful discussions, and we gratefully
    acknowledge the computational resources provided by the Vermont Advanced
    Computing Core, supported by NASA (NNX-08AO96G).

\section*{Funding}

JB has been supported by the University of Vermont and the Vermont Complex Systems Center.

\singlespacing{}




\section*{Supplementary material}

# Supporting Information

*Understanding the group dynamics and success of teams*
by Michael Klug and James P. Bagrow

## Table of Contents

## List of Figures

## S1  GitHub popularity

Since roughly 2011 GitHub has become the predominant online platform for developing and hosting open source software and other projects. Competitors such as Sourceforge have been left behind. Figure S1 shows the Google search volume for GitHub and its competitors, as gathered from Google Trends (https://www.google.com/trends/).

## S2  Number of commits per push

A "push" to GitHub can consist of multiple "commits". A commit is an individual update to the project's git repository. However, most pushes consist of only a single commit (Fig. S2). A commit can contain many or few changes to a repository, and a change eventually affecting only a small piece of a file may have taken an enormous amount of effort to achieve. This variability makes "lines of code" work metrics noisy. Despite this, they are a commonly analyzed metric.

## S3  The distribution of workload across team members

Figure S3A shows how the total workload (number of pushes) of teams grows with team size $M$, meaning larger teams are more active in total. This makes sense. The inset shows how the fraction of total work done

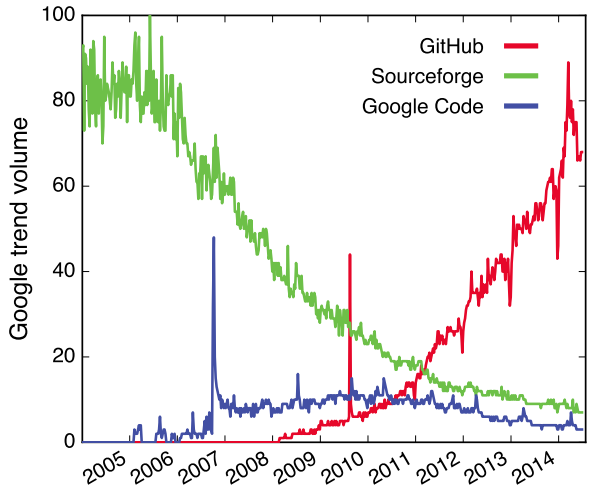

**Figure S1: GitHub is becoming the dominant platform for open development and collaboration.** Google trends (normalized) search volume over time for searches "GitHub", "Sourceforge", and "Google Code". This trend data serves as a proxy for popularity.

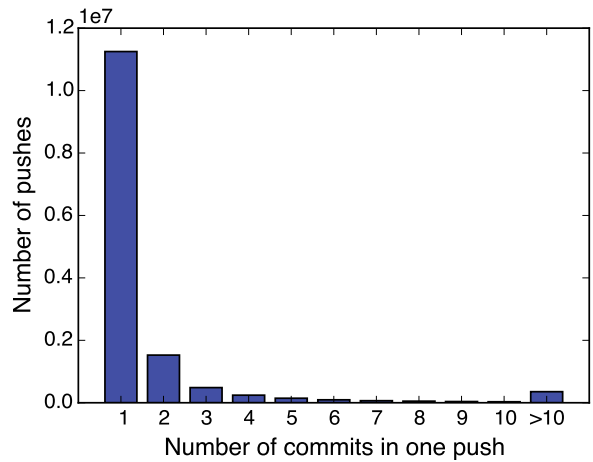

**Figure S2:** Most GitHub pushes consist of only one commit.

by the lead ($w_1/W$) decreases on average as team size grows. The lower bound $1/M$ is almost never reached, indicating that workloads remain "front-loaded" for all team sizes.

Figure S3B shows how the average workload per team member *decreases* as teams grow. The ten curves correspond to deciles of the total workload distribution. They all follow the same functional form, except possibly the highest decile which decays slightly more slowly. This indicates that the distribution of workload across team members is independent of total workload; more active teams distribute their workloads across their members in the same manner as less active teams.

## S4 Outliers in success do not skew team focus results

In Fig. S4 we reproduce the results corresponding to main text Fig. 2 but withhold the 1% highest $S$ projects. Our results do not change, indicating that the trends observed in main text Fig. 2 are not due to outliers. Note that the magnitude of the trend in figure panel D does decrease quite a bit, but the trend is still evident.

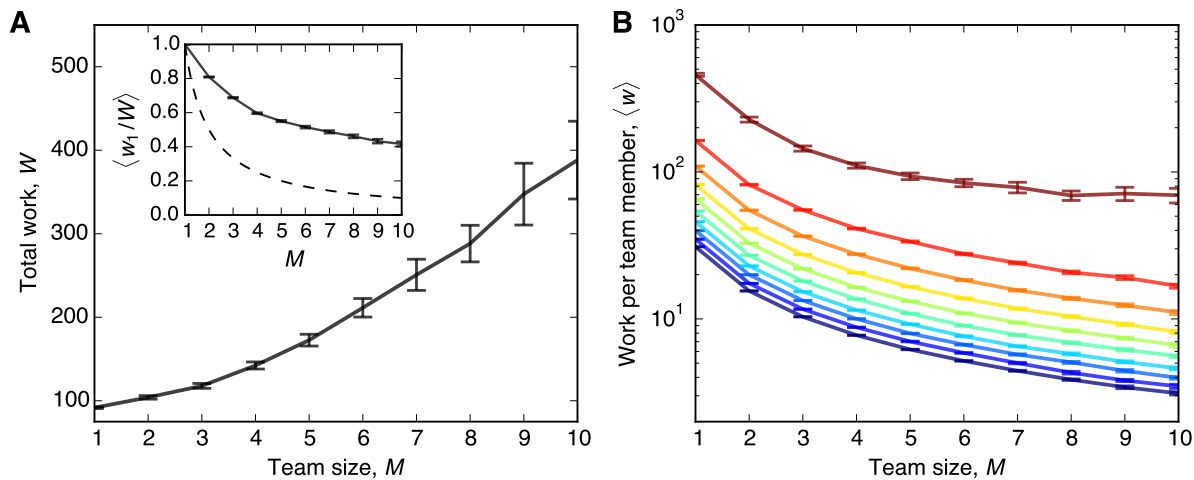

**Figure S3: Larger teams are more active but all teams distribute work the same way.** (**A**) The total work (total number of pushes) grows with team size, on average. (Inset) The average fraction of work done by the lead decreases with team size, but on average remains well above the lower bound $1/M$. (**B**) The average work per team member decreases as team size grows. The ten curves correspond to the deciles of the total work distribution. The functional form is identical for each decile, except perhaps the highest decile which decays slightly more slowly with $M$, indicating that the way in which work is distributed over a team is independent of the total activity of that team.

## S5 Experience is weakly related to team success

One of the quantities we used to understand a team is their experience $E$, defined as the average number of other teams that members of the team belong to. We found in the main text that $E$ and $S$ are significantly correlated by themselves, but that $E$ is not significant when including the other variables total team size, effective team size, diversity of experience, and number of lead members. This was shown using a linear regression model.

Here we further underscore these results. In Fig. S5A we find that top teams have significantly higher experience than average teams, but this distinction disappears for $M \geq 7$. Figure S5B shows how $E$ and $S$ relate directly, independent of other quantities. We do see a significant increase in $S$ as $E$ grows, but the change in $S$ is far smaller in magnitude than that of diversity $D$ or number of leads $L$ (shown in the main text). Moreover this effect is confounded with team size, and larger teams may even show a decrease in success as experience grows, although there are insufficient statistics to conclude a trend exists.

## S6 Linear model with secondary team contributions

In Table S1 we present the linear model corresponding to that of the main text but augmented with two additional dependent variables: $M_{\text{ext}}$, the number of users who have submitted at least one pull request to the team, and $W_{\text{ext}}$, the number of submitted pull requests. These variables are both significantly correlated with team success $S$. Including them in the linear regression model did not change the significances of any other variables although it did alter the regression coefficients.

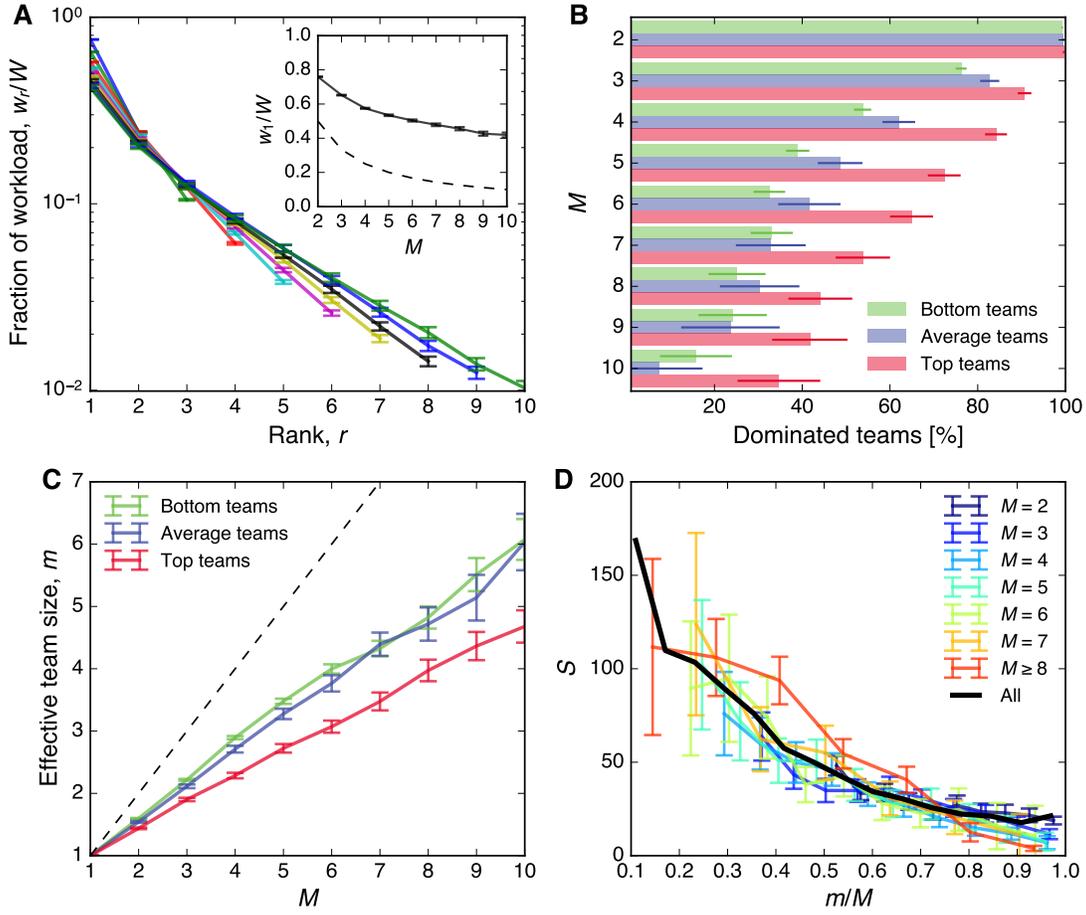

**Figure S4:** Teams are focused, and top teams are more focused than average teams of the same size. As per main text Fig. 2 but calculations now exclude outliers in $S$ (above the 99th percentile). Our results do not change after removing outliers.

| Variable $x$ | Coefficient $\beta_x$ | $p$ |
|---:|---:|---|
| constant $\alpha$ | $2.416 \times 10^{-15} \pm 0.004489$ | 1 |
| Team size, $M$ | $-0.1165 \pm 0.012856$ | $< 10^{-69}$ |
| Eff. team size, $m$ | $0.0372 \pm 0.011107$ | $< 10^{-10}$ |
| Total work, $W$ | $0.0176 \pm 0.004563$ | $< 10^{-13}$ |
| Experience, $E$ | $-0.0032 \pm 0.004520$ | 0.16267 |
| Diversity, $D$ | $0.0072 \pm 0.005776$ | 0.01506 |
| Num. of leads, $L$ | $0.0534 \pm 0.006371$ | $< 10^{-59}$ |
| Age, $T$ | $0.0916 \pm 0.004563$ | 0 |
| Num. pull requesters, $M_{\text{ext}}$ | $0.4684 \pm 0.005480$ | 0 |
| Num. pull requests, $W_{\text{ext}}$ | $-0.0515 \pm 0.005354$ | $< 10^{-78}$ |

**Table S1:** OLS regression model on team success, $S = \alpha + \beta_M M + \beta_m m + \beta_W W + \beta_E E + \beta_D D + \beta_L L + \beta_T T + \beta_{M_{\text{ext}}} M_{\text{ext}} + \beta_{W_{\text{ext}}} W_{\text{ext}}$. Outliers (above the 99th percentile in $S$) were filtered out to ensure they do not skew the model.

4

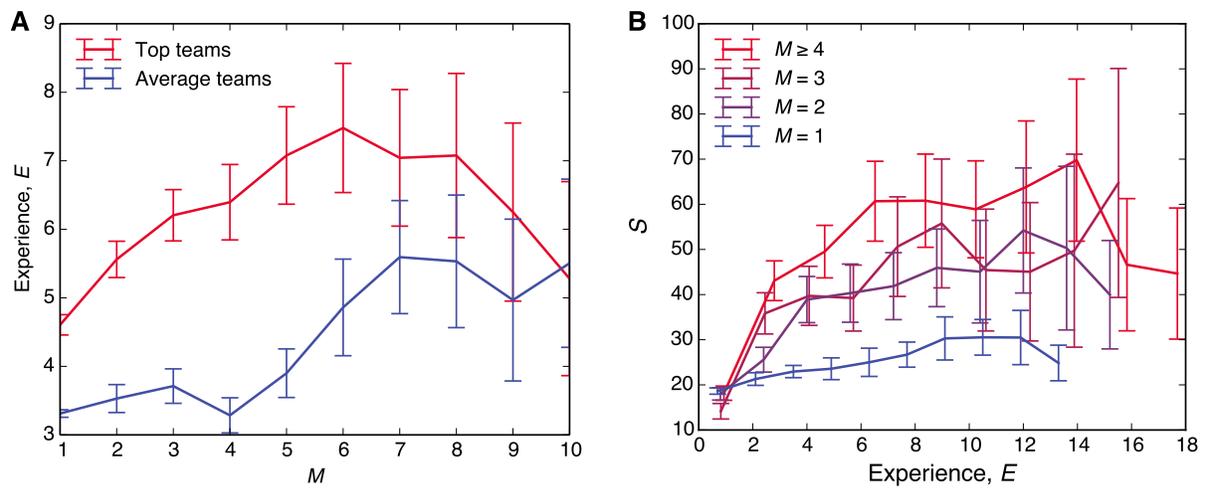

**Figure S5: Experience weakly correlates with success.** (**A**) Top teams have higher experience than average teams, but only for team sizes $M < 7$. (**B**) As experience grows, success grows on average, but the change in $S$ is quite weak, compared to that of diversity and number of lead members (main text). Linear models also indicated that the effects of $E$ are entirely due to the other quantities.



\end{document}